# Simulation Study of the Effects of Polymer Network Dynamics and Mesh Confinement on the Diffusion and Structural Relaxation of Penetrants

Tsai-Wei Lin[1,3], Baicheng Mei[2,3], Kenneth S. Schweizer[1,2,3], and Charles E. Sing[1,3]

[1]Department of Chemical and Biomolecular Engineering, University of Illinois at Urbana-Champaign

[2]Department of Materials Science and Engineering, University of Illinois at Urbana-Champaign

[3]Materials Research Laboratory, University of Illinois at Urbana-Champaign

**Abstract**

The diffusion of small molecular penetrants through polymeric materials represents an important fundamental problem, relevant to the design of materials for applications such as coatings and membranes. Polymer networks hold promise in these applications, because dramatic differences in molecular diffusion can result from subtle changes in the network structure. In this paper, we use molecular simulation to understand the role that crosslinked network polymers have in governing the molecular motion of penetrants. By considering the local, activated alpha relaxation time of the penetrant and its long-time diffusive dynamics, we can determine the relative importance of activated glassy dynamics on penetrants at the segmental scale versus entropic mesh confinement on penetrant diffusion. We vary several parameters, such as the crosslinking density, temperature, and penetrant size, to show that crosslinks primarily affect molecular diffusion through modification of the matrix glass transition, with local penetrant hopping at least partially coupled to the segmental relaxation of the polymer network. This coupling is very sensitive to the local activated segmental dynamics of the surrounding matrix, and we also show that penetrant transport is affected by dynamic heterogeneity at low temperatures. To contrast, only at high temperatures and for large penetrants or when the dynamic heterogeneity effect is weak does the effect of mesh confinement become significant, even though penetrant diffusion more broadly empirically follows similar trends as established models of mesh confinement-based transport.

**Introduction**

The diffusive transport of molecular species through polymers is an important fundamental problem, with the motion of small 'penetrants' being integral to the design of polymer materials for a variety of applications. For example, barrier coatings,[1–3] drug delivery vehicles,[4,5] and self-healing microcapsules[1,2,6] may be engineered to impede the transport of penetrants, while membranes are often designed to selectively separate or filter specific small penetrants (e.g. gas,[7–10] water,[11–13] or organic molecules[14,15]). Membrane separations represent an especially important materials design challenge,[7,15] due to significant interest in finding promising alternatives to distillation for chemical separations. One strategy is to impose control over pore size in materials such as zeolites,[16] metal organic frameworks,[17,18] and covalent organic frameworks;[19] however, these materials have issues with mechanical properties and scalability.[16–19] If the molecular penetrants are especially small, such as in gas separations, then glassy amorphous polymers can be another option for membranes.[8–10] These materials are more mechanically robust, but despite important insights from atomistic simulations and phenomenological models,[20–23] they are limited to small gas molecules and do not have the same size-selectivity as more precise matrix structures.[8–10]

Highly-crosslinked networks have recently emerged as a promising option for the selective transport of small molecules,[24–26] exploiting the sensitivity of penetrant diffusion to the near-$T_g$ coupling between the structural relaxation of the dense molecular mesh and activated hopping processes. This is motivated by experimental support for the premise that, despite significant molecular disorder, dense polymer networks can discern small changes in penetrant size and interactions.[25] Design of crosslinked networks for selective transport requires an understanding of how the molecular-scale dynamics of penetrant and matrix relaxation contribute to large-scale diffusive motion. Aspects of this problem have been studied in the literature. For example, there has been extensive work on studying particle transport through polymer solutions or high-temperature networks,[27–33] which is understood to be governed by a confinement parameter $C = d/a_x$ that relates the penetrant size $d$ to the size of the network mesh $a_x$. This parameter quantifies the extent to which the surrounding polymer matrix sterically 'traps' the particle, such that the particle diffusion is an activated process where the network strands must entropically stretch to allow the particle to hop out of its location within the mesh.[27] Several expressions for the diffusion constant have been considered in the literature, such as the expression proposed by Cai, *et al.*:[27]

$$D_p \sim \frac{d^2}{\tau} C^{-1} \exp(-bC^2) = \frac{d^2}{\tau} X \qquad (1)$$

This relates the diffusion constant $D_p$ to the confinement parameter $C$, a length scale characterized by the penetrant size $d$, a time scale $\tau$ that Cai, *et al.* identify as the Rouse time of the network strand, and a constant $b$ of order unity.[27] While there are other candidate models,[30] these still exhibit a general form given in the second equivalence of **Equation 1** as the product of two factors; a quantity proportional to the elementary particle diffusion $d^2/\tau$ and a factor $X$ dictated by the network mesh and related to $C$ (in the case above, $X = \exp(-bC^2)/C$). Simulations of large penetrants in rubbery networks are consistent with this physical picture;[28,29,34] however, the choice of $\tau$ can become non-trivial as temperature is lowered for polymer melts and/or when the glass transition is approached.[35] For example, in uncrosslinked systems activated transport of penetrants becomes dominated by molecular caging and correlated matrix elasticity effects near the glass transition.[35–38]

For tight and supercooled near-$T_g$ networks, the physical relationship in **Equation 1** is complicated by recent experimental, simulation, and theoretical investigations by the authors and collaborators demonstrating that the structural relaxation in polymer networks is strongly dependent on the extent of crosslinking, manifesting as a monotonic increase in the material $T_g$ with crosslink density.[39] In this case, the time scale $\tau$ in **Equation 1** is expected to depend on molecular penetrant caging, which itself will be sensitive to the confinement parameter $C$ due to its relationship with the crosslink density of the polymer network. Crosslinks could thus impact penetrant motion both by affecting the structural relaxation of the surrounding network as well as providing a confining mesh that obstructs penetrant motion. Our recent work demonstrated using a combination of theory, experiment, and simulation that there exists a degeneracy of interpretations for crosslink effects on penetrant transport between these two mechanisms, noting the need to further clarify the molecular origin of this observation.[40]

In this article, we use computer simulation to study the diffusion and structural relaxation of penetrants in dense, highly-crosslinked polymer networks in the computationally accessible weakly-supercooled regime. Simulation will allow us to separate out the different ways that crosslinks contribute to penetrant diffusion, by quantifying both the overall diffusion constant $D_p$ *and* the characteristic time scale of penetrant hopping $\tau$ in **Equation 1**, which we identify with a penetrant alpha relaxation time $\tau_{\alpha,p}$ determined from an appropriate time correlation function as discussed below. By relating particle hopping events to long-time diffusive motion, we can study how these properties are affected by both crosslinking fraction and temperature. We determine that (1) supercooled networks exhibit strong coupling between the network segmental relaxation and penetrant hopping, with sensitivity to the crosslinking-dependent $T_g$, (2) increasing temperature can lead to slower diffusion at a given $T_g/T$ due to the dependence of $T_g$ on crosslink density, but this is mostly attributed to a non-trivial $T$ dependence of how local polymer relaxations couple to penetrant motion, (3) the coupling between penetrant motion and polymer relaxation processes is dependent on penetrant size $d$, and (4) penetrant hopping is correlated with long-time diffusion at several values of $T_g/T$ near $T = T_g$, but the mesh confinement only becomes apparent at high $T$ and for large penetrants. This all leads to the overall insight that, under typical conditions for polymer networks and molecular penetrants, crosslinking primarily affects penetrant transport by changing the effective $T_g$ and segmental relaxation time of the surrounding matrix and only *secondarily* through mesh confinement. Serendipitously, this still manifests in a similar relationship between the diffusion constant $D_p$ and $C$ that is consistent with the form in **Equation 1**.[27,30] These molecular-level insights into penetrant diffusion and hopping clarify the fundamental mechanisms governing penetrant transport and help identify ways in which both temperature and network architecture can be used to engineer selectivity into a promising class of membrane materials.

We also briefly discuss at a high level how the theoretical results in our companion paper[41] qualitatively compare to our present simulations. The goal is not to quantitatively use simulation to test the theory, especially since the polymer and penetrant models, and range of temperatures explored, are not the same. Rather, the present work and the theoretical companion paper are meant to be complementary with regards to addressing the scientific problems of interest.

**Simulation Methods**

We use coarse-grained molecular dynamics simulations (MD) to model the diffusion of spherical penetrants in crosslinked polymer networks,[42] using well-established methods for studying polymers near the glass transition.[43–46] Our system is initially composed of $N_c = 20$ linear chains with $N_m = 30$ beads each with diameter $\sigma$, modeled as standard semiflexible chains.

These beads are placed in a cubic box with periodic boundary conditions in three dimensions, along with $N_\text{p}$ spherical penetrants of diameter $\tilde{d}$. The number of penetrants $N_\text{p}$ depends on $\tilde{d}$, and is chosen so that the volume fraction of the penetrant $\phi_p = \frac{\pi d^3 N_p}{6V}$ is kept below $\phi_p = 0.01$ to ensure that the addition of penetrants will not affect network dynamics and to prevent significant interactions between penetrant molecules.

Standard dimensionless simulation quantities are employed,[42,47] and denoted with tildes (e.g. $\tilde{T} = T/T^*$). Quantities are rendered dimensionless by several characteristic values: lengths are related to the bead diameter $\sigma$, energies are related to the thermal energy at a reference temperature $k_B T^*$, and times are related to a time scale $\tau^* = \sqrt{m\sigma^2/k_B T^*}$, where $m$ is the monomer mass. The characteristic temperature $T^*$ corresponds to 485K, which is determined from parametrization against experiments,[39] using crosslinked poly(n-butyl acrylate) (PnBA) as a representative polymer that has been studied by our experimental collaborators (see details in ref. 35).

Our MD simulations consider standard potentials for coarse-grained bead-spring polymers,[48] including a Lennard-Jones potential $\tilde{U}_\text{LJ}$ between all non-bonded beads, and both bonding $\tilde{U}_\text{B}$ and bending $\tilde{U}_\theta$ potentials to model each semiflexible bead-spring chain. The overall energy is composed of a sum of these contributions:

$$\tilde{U} = \tilde{U}_\text{B} + \tilde{U}_\theta + \tilde{U}_\text{LJ}$$
$$= \sum_{i,j>i}^{N_{tot,n}} \tilde{u}_{\text{B},ij} + \sum_{i,j>i,k>i,j}^{N_{tot,n}} \tilde{u}_{\theta,ijk} + \sum_{\alpha=n,p} \sum_{\beta=n,p} \sum_{i}^{N_{tot,\alpha}} \sum_{j>i}^{N_{tot,\beta}} \tilde{u}_{\text{LJ},\alpha\beta,ij} \quad (2)$$

Here the total system energy is written in terms of independent pairwise contributions $\tilde{u}_{\text{B},\alpha,i}$, $\tilde{u}_{\theta,\alpha,i}$, and $\tilde{u}_{\text{LJ},\alpha\beta,ij}$. Bonded monomers interact through harmonic bonding potential:

$$\tilde{u}_{\text{B},ij} = \frac{\tilde{k}}{2}(\tilde{r}_{ij} - 1)^2 \Theta_{ij} \tag{3}$$

A large spring constant $\tilde{k} = 2000$ is adopted to enforce the equilibrium distance $\tilde{r}_{ij} = 1$ between bonded beads $i$ and $j$. The factor $\Theta_{ij}$ determines the connectivity between monomer pairs, with $\Theta_{ij} = 1$ being connected and $\Theta_{ij} = 0$ being disconnected; these values depend on the specific network that is formed. A bending energy is similarly introduced to account for chain stiffness:

$$\tilde{u}_{\theta,ijk} = \tilde{k}_\theta [1 - \cos\theta_{ijk}] \Theta_{ijk} \tag{4}$$

Here, the bond angle is between three adjacent beads $i, j$, and $k$ whose connectivity is determined by a similar factor $\Theta_{ijk}$ to the one used for the bonding potential. The bending constant $\tilde{k}_\theta = 1.52$ is selected to reflect the experimental Kuhn length of PnBA.[49–52] Further details of the parametrization are given in the literature.[39] A Lennard-Jones (LJ) potential is used to describe all non-bonded interactions between particles $i$ and $j$ on species $\alpha$ and $\beta$:

$$\tilde{u}_{\text{LJ},\alpha\beta,ij} = \begin{cases} 4\tilde{\epsilon}_{\alpha\beta}\left[\left(\frac{\tilde{d}_\alpha + \tilde{d}_\beta}{2\tilde{r}_{\alpha\beta}}\right)^{12} - \left(\frac{\tilde{d}_\alpha + \tilde{d}_\beta}{2\tilde{r}_{\alpha\beta}}\right)^6\right], & \tilde{r}_{\alpha\beta} < \tilde{r}_{\text{cut}} = 2.5 \times \frac{\tilde{d}_\alpha + \tilde{d}_\beta}{2} \\ 0, & \text{otherwise} \end{cases} \tag{5}$$

Here, $\alpha, \beta \in \{n, p\}$, denotes the species as either network (n) or penetrant (p), with n denoting the monomer bead in the network and p denotes the penetrant. In the overall summation, $N_{tot,n} = N_C N_m$ is the total number of network beads and $N_{tot,p} = N_p$ is the total number of penetrant beads. The monomer size $\tilde{d}_n = 1$ is always unity (i.e., $d_n = \sigma$) and the penetrant size $\tilde{d}_p \equiv \tilde{d}$ is an important parameter for this study. In this study, we do not consider specific penetrant-polymer attractions, so we set $\tilde{\epsilon}_{nn} = \tilde{\epsilon}_{np} = \tilde{\epsilon}_{pp} = 1$.

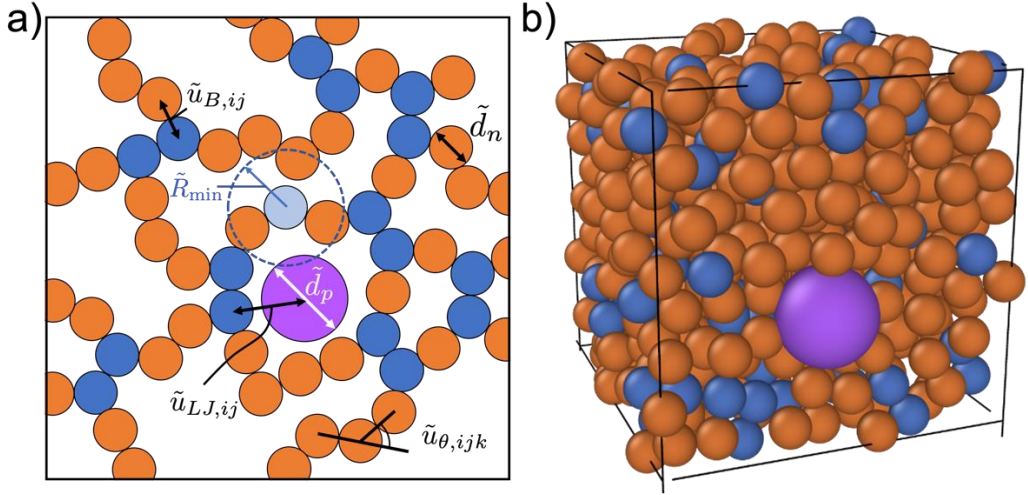

**Figure 1**. (a) Schematic illustrating the setup of our simulation. Monomer beads (orange) and crosslinker beads (blue) interact via bonding ($\tilde{u}_{B,ij}$) and angle ($\tilde{u}_{\theta,ijk}$) potentials, and can interact with each other and with the penetrant through Lennard-Jones potentials ($\tilde{u}_{LJ,ij}$). Monomer and crosslinker beads have a diameter $\tilde{d}_n = 1$ and the penetrant has a diameter $\tilde{d}_p$. Crosslinking occurs during an initial simulation step where 'reactive' beads (light blue) will form bonds with regular monomers within a cutoff distance $\tilde{R}_{\min}$. (b) Snapshot of a typical simulation, for $d/\sigma = 2.0$ and $f_{cross} = 0.11$.

The system is first equilibrated at $\tilde{T} = 1$, $\tilde{P} = 0$, and then networks are prepared by crosslinking the linear chains with reactive beads randomly distributed along the chain.[39,53,54] The total number of reactive beads is $N_r = f_r N_m N_c$, where $f_r$ is the fraction of reactive beads which tunes the crosslink density of networks. If the distance between a reactive bead and a free bead (orange beads in **Figure 1**) is within $\tilde{R}_{\min} = 1.1$, a new permanent bond will be formed given an assigned probability. Once a reactive bead forms a bond with a free bead, both the reactive bead and the free bead are labeled as crosslink beads (dark blue beads in **Figure 1**) and these two beads represent one crosslink 'molecule', in a way that reflects the specific chemistry used to synthesize PnBA networks,[25] where the crosslinker molecular weight is roughly twice of the molecular weight of nBA. Crosslinking reactions were turned off once every reactive bead has formed a new bond with another free monomer (maximum number of possible bonds has been reached). In the end, $N_r$ reactive beads have reacted with the $N_r$ free beads and resulting in $2N_r$ crosslink beads associated with $N_r$ crosslink molecules. The crosslink density, $f_{cross}$, of networks is defined as:[39]

$$f_{cross} = \frac{n_{crosslink}}{n_{crosslink} + n_{monomer}} = \frac{N_r}{N_m N_c - N_r} \tag{6}$$

Here, $n_{crosslink}$ and $n_{monomer}$ are the number of crosslink molecules and nBA monomers, respectively. Four values of crosslink fraction $f_{cross}$ are considered: 0.11, 0.25, 0.36, and 0.5.

After crosslinking, the system is cooled to a target temperature at a cooling rate $\tilde{\Gamma} = 8.3 \times 10^{-6}$ (corresponding to $\Gamma = 1.25 \times 10^9$ K/s in experimental units) and further equilibrated at constant $\tilde{P} = 0$. Another short NPT run was performed and the mean volume $V$ is measured. We then switch to a NVT ensemble by setting the system volume to the mean volume $V$ and equilibrate the system before final production run at NVT. All simulations are performed in LAMMPS[55] with a standard Nosé-Hoover thermostat and extended ensemble barostat,[42,56,57] and all observables were averaged over five independent simulation trajectories.

The polymer network is characterized by its mesh size ($\tilde{a}_x$), Kuhn segment alpha relaxation time ($\tilde{\tau}_{\alpha,K}$), and glass transition temperature $T_g$. The mesh size of networks at different crosslink densities is defined as the averaged distance between two adjacent crosslink beads on the same chain, and is tabulated in **Table S1**. The value of $\tilde{\tau}_{\alpha,K}$ at each temperature and crosslink density was defined as the time where the temporal autocorrelation function of a Kuhn monomer vector, $C_\lambda(\tilde{t}) = \langle P_2(\tilde{r}_3(\tilde{t}) \cdot \tilde{r}_3(0))\rangle$ decays to $C_\lambda(\tilde{t}_{max}) + (1 - C_\lambda(\tilde{t}_{max}))/e$, where $C_\lambda(\tilde{t}_{max})$ represents the plateau value at the long-time limit $\tilde{t}_{max}$.[58–62] Here, $P_2$ is the second Legendre polynomial and $\tilde{r}_3$ is a vector between two beads that are 3 bonds apart which reflects the choice of coarse-grained bead relate to Kuhn segment of PnBA. $T_g$ is then defined as when the Kuhn segment alpha relaxation time is $\tilde{\tau}_{\alpha,K}(\tilde{T}_g) = 10^5$. More information can be found in our previous work.[39]

We characterize the dynamics of the penetrant motion by its alpha relaxation time, $\tilde{\tau}_{\alpha,p}$, and its diffusion coefficient $D_p$. To calculate $\tilde{\tau}_{\alpha,p}$, we use the self-intermediate scattering function (ISF) of the penetrant, which is given by:[59]

$$F_s(\tilde{\boldsymbol{q}}, \tilde{t}) = \frac{1}{N_p} \sum_j^{N_p} \left\langle \exp\left[-i\tilde{\boldsymbol{q}} \cdot \left(\tilde{\boldsymbol{r}}_{pj}(\tilde{t}) - \tilde{\boldsymbol{r}}_{pj}(0)\right)\right]\right\rangle \quad (7)$$

where $\tilde{\boldsymbol{r}}_{pj}$ is the position of the $j^{\text{th}}$ penetrant at time $\tilde{t}$, and $|\tilde{\boldsymbol{q}}|$ is set to the position of the first peak of the static structure factor for the networks, corresponding to $|\tilde{\boldsymbol{q}}| = 7.2$. The value of $\tilde{\tau}_{\alpha,p}$, is defined as the time required for $F_s(\tilde{\boldsymbol{q}}, \tilde{t})$ to decay to either $1/e$ or $0.1$ from its initial ($\tilde{t} = 0$) value of unity. We characterize the mean-squared displacement (MSD) as a function of time: $\text{MSD}(\tilde{t}) = \langle \tilde{r}_p^2(\tilde{t}) \rangle = \frac{1}{N_p} \sum_j^{N_p} \left\langle \left(\tilde{\boldsymbol{r}}_{pj}(\tilde{t}) - \tilde{\boldsymbol{r}}_{pj}(0)\right)^2 \right\rangle$. The diffusion coefficient of penetrant is then obtained via the Einstein relation in the diffusive regime, $D_p = \lim_{\tilde{t} \to \infty} \frac{1}{6\tilde{t}} \langle \tilde{r}_p^2(\tilde{t}) \rangle$.[63]

**Results and Discussion**

*Penetrant Hopping and Transport in Dense Networks*

We first characterize the diffusive motion of penetrants in polymer networks by studying the mean square displacement, $\langle \tilde{r}_p^2(\tilde{t}) \rangle$, of both small ($d/\sigma = 1.0$) and large ($d/\sigma = 2.0$) penetrants. These two particle sizes are representative of experimental systems used in previous work by some of the authors;[26] the network polymer PnBA is coarse-grained so that $\sigma = 0.573$nm, and hence $d/\sigma \sim 1 - 2$ spans the typical size of molecular penetrants.[26] This choice also corresponds to the general length scale of the mesh size in the literature,[25,64] which here ranges size from $\tilde{a}_x \sim 1.3 - 2.5$ (see **Table S1**) such that the size ratio range spans $C = d/a_x = 0.4 - 1.5$. The MSD versus time $\tilde{t}$ is plotted for both particle sizes in Figures 2a and 2b for $d/\sigma = 1.0$ and $d/\sigma = 2.0$, respectively, varying both the crosslink density $f_{cross}$ and temperature. These plots show typical features characteristic of molecular diffusion. At short times, the particle exhibits ballistic motion such that the MSD scales as $\langle \tilde{r}_p^2(\tilde{t}) \rangle \sim \tilde{t}^2$; however, this regime gives way to a subdiffusive regime that we attribute to molecular caging by neighboring network monomers. At long times, the particle can overcome the kinetic constraints imposed by these cages, along with any constraints due to the network mesh, and undergo Fickian diffusion where $\langle \tilde{r}_p^2(\tilde{t}) \rangle \sim \tilde{t}^1$. These general features of diffusive motion are observed for both penetrant sizes and all crosslink densities and temperatures, however the specifics of penetrant caging and diffusive motion will be affected by these parameters.

The onset of the subdiffusive regime following the initial, ballistic penetrant motion provides insight into the time scales of caging, and large differences in the persistence of this regime are observed as $f_{cross}$, $T$, and $d/\sigma$ are varied. To characterize this regime, we plot in **Figure S1** the slope of the $\log\langle \tilde{r}_p^2(\tilde{t}) \rangle$ versus $\log \tilde{t}$ curve on a log-log plot ($\beta = d\log\langle \tilde{r}_p^2(\tilde{t}) \rangle / d\log \tilde{t}$) as a function of time $\tilde{t}$. The minimum value of $\beta$ is an estimate for the time $\tilde{t}_\beta$ at which the penetrant is maximally subdiffusive or caged, and the value of $\langle \tilde{r}_p^2(\tilde{t}_\beta) \rangle^{1/2}$ correspondingly quantifies a transient localization length. A representative quantity is indicated in **Figures 2a** and **2b** with an open square, showing the caging onset for $T = 300$ and $f_{cross} = 0.11$. We only indicate a single representative value for each particle size, because the displacement $\langle \tilde{r}_p^2(\tilde{t}_\beta) \rangle^{1/2}$ is only weakly dependent on temperature and crosslink densities, and is far smaller than the length scale of the penetrant diameter ($\langle \tilde{r}_p^2(\tilde{t}_\beta) \rangle^{1/2} \sim 0.22$ for $d/\sigma = 1$ and $\langle \tilde{r}_p^2(\tilde{t}_\beta) \rangle^{1/2} \sim 0.32$ for $d/\sigma = 2$). This insensitivity leads us to interpret this onset of caging as tied to the structural correlations of the surrounding melt beads.

For both small ($d/\sigma = 1$) and large ($d/\sigma = 2$) penetrants, the value of $\beta \to 1$ at long times corresponds to Fickian behavior, however the time that it takes to reach this limit increases with decreasing temperature (**Figure 2** and **Figure S1**). This is the expected result of having less thermal energy to overcome a dynamic caging barrier,[24,65] especially as our simulations approach the $T_g$. Similarly, an increase of crosslink density $f_{cross}$ monotonically increases the length of this subdiffusive regime and the minimum value of $\beta$ concomitantly decreases (**Figure S1**) so that the MSD nearly exhibits a transient plateau. In principle, this is due to some combination of (1) an increase in the caging barrier because of the concomitant increase in the segmental relaxation time (cage lifetime) due to crosslinks[39] and (2) the impact of mesh confinement impeding penetrant motion.[27,28,30] In this paper, a major result will be to show that the first interpretation is the dominant determinant of penetrant mobility for the models studied, especially as $T_g$ is approached. This is apparent already by comparing the features in the MSD versus $\tilde{t}$ to the mesh size $\tilde{a}_x$, which are tabulated in **Table S1** for several different values of $f_{cross}$ and indicated in **Figures 2** and **S1** as open circles. This length scale is often only accessed in or near the Fickian regime, suggesting that any effect of the geometric mesh on the subdiffusive transport of the penetrant in these simulations is subtle and not readily apparent from the MSD versus $\tilde{t}$ plot alone. Despite this observation, it is apparent that both the temperature and crosslink density effects still depend on the penetrant size, with more pronounced changes in the subdiffusive regime for the larger ($d/\sigma = 2$) penetrants than the smaller ($d/\sigma = 1$) penetrants as $T$ and $f_{cross}$ are varied.

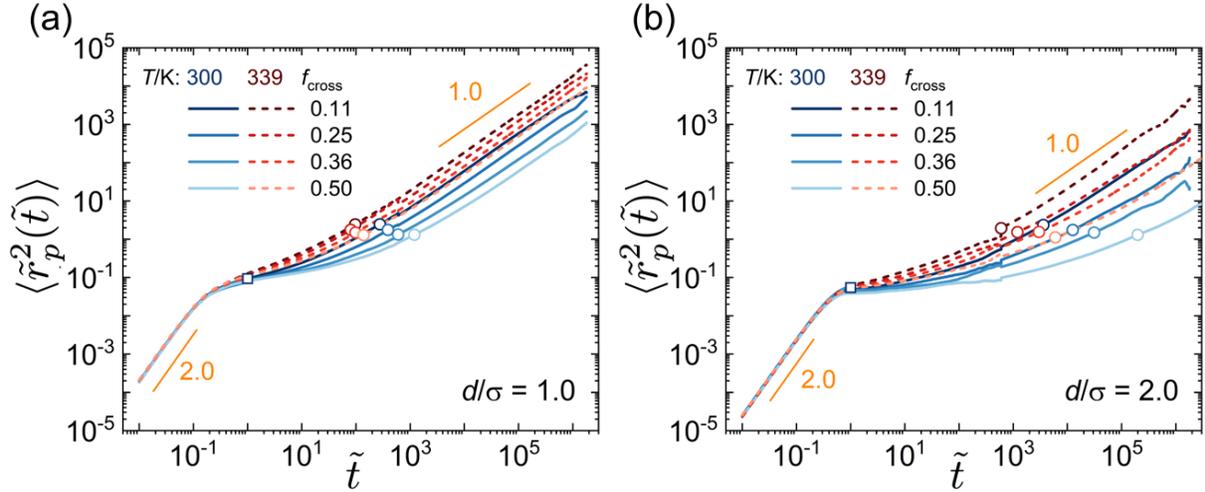

**Figure 2**. Mean square displacement $\langle \tilde{r}_p^{\,2}(\tilde{t}) \rangle$ versus time $\tilde{t}$ of a penetrant particle with sizes (a) $d/\sigma = 1.0$ and (b) $d/\sigma = 2.0$, for several values of $T$ and $f_{cross}$. Open squares denote a representative location for the onset of caging or transient localization, defined as the minimum slope $\beta$ on this plot (see **Figure S1**) and chosen for $T = 300$K and $f_{cross} = 0.11$. The analogous point for other values of $T$ and $f_{cross}$ would be almost identical. The open circles denote the length scale of the polymer network mesh size, as described in **Table S1**. The long-time data Fickian region, where the slope of this plot $\beta = 1$, is used to determine the penetrant diffusion constant $\widetilde{D}_p$.

The MSD provides a molecular measure of penetrant transport, with the long-time Fickian regime allowing determination of the macroscopic diffusion coefficient $\widetilde{D}_p$ of the penetrant. However, this quantity ($\widetilde{D}_p$) is determined by, in principle, the cumulative effect of both local particle motions as well as any larger length-scale effects of the network mesh confinement on transport. To deconvolute these contributions, we also consider the time evolution of the self-ISF $F_s(\widetilde{\boldsymbol{q}}, \tilde{t})$ in **Figure 3**. Here we select a value $|\widetilde{\boldsymbol{q}}| = 7.2$ that is chosen to reflect the local structural cage of the polymer network and *not* the length scale of the network mesh. The curves shown in **Figure 3** thus primarily account for the dynamics of local relaxation and hopping, and quantify the extent to which the penetrant remains within its original cage as a function of time. The first part of this decay is independent of temperature and crosslinking; this corresponds to the initial ballistic motion of the penetrant observed in **Figure 2,** and may include contributions from partial fast relaxations due to a distribution of network alpha times associated with dynamic heterogeneity.[66] There are some subtle differences between the different penetrant sizes ($d/\sigma = 1.0$ vs. $2.0$) that we attribute to the different masses of the penetrants and is also apparent in Figure 2.

At longer times, the decay of $F_s(\tilde{\mathbf{q}}, \tilde{t})$ becomes strongly dependent on temperature, crosslinking, and penetrant size in a manner consistent with the diffusive motion in **Figure 2**. At high temperatures, the smaller penetrant (**Figure 3a**) exhibits a non-exponential, long time decay of $F_s(\tilde{\mathbf{q}}, \tilde{t})$ that we attribute to an activated, but relatively rapid, hopping process for penetrant relaxation. As crosslinking increases, this long-time decay extends further, and its non-exponential character becomes more pronounced. This trend is exacerbated by going to lower temperatures. We find there is a significant slowing down of penetrant motion even at intermediate times; this occurs around $\tilde{t} \sim 1$, which is roughly when the minimum in the apparent exponent $\beta$ occurs for the MSD curves (in **Figure 2** and **S1**) and is consistent with the interpretation of this as a state of maximal transient penetrant caging. For larger penetrants ($d/\sigma = 2.0$, **Figure 3b**) this slowing down leads to the near transient arrest of particle motion, again at $\tilde{t} \sim 1$, as evidenced by a plateau in $F_s(\tilde{\mathbf{q}}, \tilde{t})$ that only begins to relax at significantly longer times as the temperature is decreased or crosslinking fraction is increased. In both cases, we can attribute this arrest to the coupling of penetrant hopping to the slow segmental relaxation dynamics of the surrounding polymer segments, as $F_s(\mathbf{q}, \tilde{t})$ is a local measure of relaxation that does not directly probe the length scales relevant for mesh confinement. We note the penetrant hopping should also be significantly affected by the crosslinking degree, although indirectly, since crosslinking has strong influence on the slow segmental relaxation dynamics of the surrounding polymer segments.[39]

The trends seen in the intermediate scattering function plots in **Figure 3** are largely consistent with those of the MSD plots in **Figure 2**, however we note several important differences that we will use to understand the mechanisms of penetrant motion in polymer networks. First, we note the similarity of the MSD versus time plots in **Figure 2a** for low temperatures and crosslink density ($T = 300$K and $f_{cross} = 0.11$) and the high temperatures and crosslink density ($T = 339$K and $f_{cross} = 0.36$); these curves almost overlap, and exhibit essentially the same diffusive properties. Yet, the shapes of these curves in **Figure 3a** for penetrant ISF are distinctly different, with the characteristics of local caging more apparent in the lower temperature curve. In addition, there is a distinct difference in the subdiffusive regimes seen in the MSD plots versus $F_s(\tilde{q}, \tilde{t})$, which are much more pronounced. This is contrast to normal glass-forming liquids, for which the same supercooling degree exhibits a weaker or shorter caging regime than for the MSD that probes diffusion compared to $F_s(\tilde{q}, \tilde{t})$ which probes structural relaxation.[66–73] These disparities between the MSD versus time and $F_s(\tilde{q}, \tilde{t})$ suggest that mesh confinement may contribute meaningfully to the transport of penetrants in tight networks, particularly for larger penetrants, and we seek to refine our physical understanding of this complicated interplay of local and long-time penetrant and network dynamics.

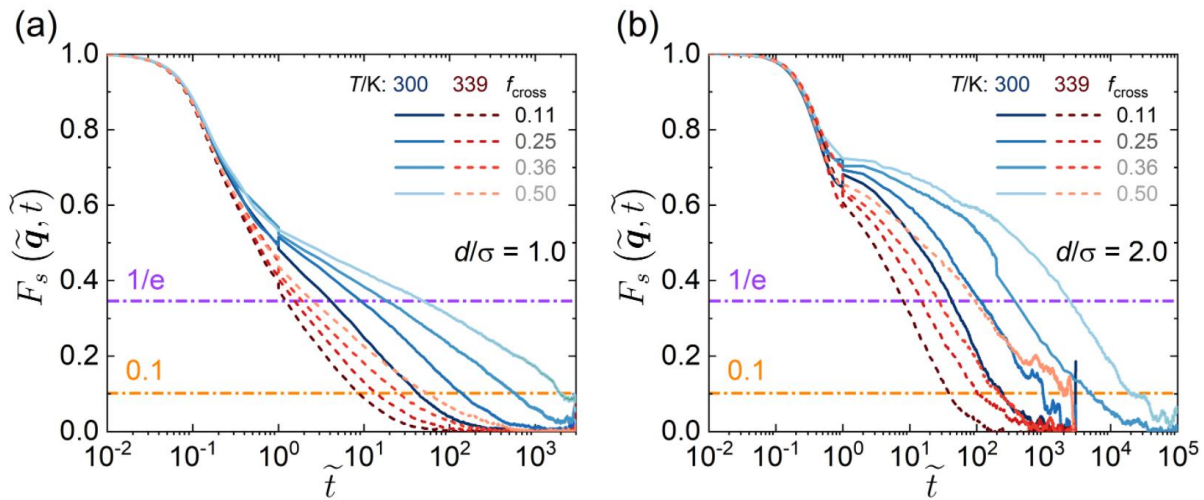

**Figure 3**. Intermediate scattering function $F_s(\tilde{q}, \tilde{t})$ of the penetrant particle with (a) $d/\sigma = 1.0$ and (b) $d/\sigma = 2.0$ with $|\tilde{q}| = 7.2$ at $T = 300$K (solid, blue) and 339K (dashed, red) for a variety of crosslink densities $f_{cross}$ as a function of time $\tilde{t}$. The orange and purple horizontal lines define the penetrant alpha time, with the criteria $F_s(\tilde{q}, \tilde{\tau}_{\alpha,p}) = 0.1$ and $F_s(\tilde{q}, \tilde{\tau}_{\alpha,p'}) = 1/e$, respectively. The former criteria ($\tilde{\tau}_{\alpha,p}$) is more often used in this paper, because it includes the longer time tails observed in these functions that reflect a subpopulation of penetrants in long-lived cages. The latter criteria ($\tilde{\tau}_{\alpha,p'}$) may more closely relate to the average hopping time that is predicted in the theory developed in our companion paper,[41] which does not account for dynamic heterogeneity and a distribution of penetrant hopping times.

*Comparing the Role of Particle Size, Temperature, and Crosslinking on the Hopping and Diffusive Processes in Dense Networks*

To quantify the relationship between local penetrant relaxation and diffusive motion, we extract the alpha time of the penetrant $\tilde{\tau}_{\alpha,p}$ from the ISF and the diffusion constant $\widetilde{D}_p$ from the MSD. As described in the methods section, the diffusion constant is obtained from the long-time MSD curve via the expression $D_p = \lim_{t \to \infty} \frac{1}{6\tilde{t}} \langle \tilde{r}_p^2(\tilde{t}) \rangle$. The alpha time of the penetrant $\tilde{\tau}_{\alpha,p}$ is set from the decay of $F_s(\tilde{q}, \tilde{t})$, and quantifies the time it takes to reach one of two possible criteria: $F_s(\tilde{q}, \tilde{\tau}_{\alpha,p}) = 0.1$ and $F_s(\tilde{q}, \tilde{\tau}_{\alpha,p'}) = 1/e$. While the specific criterion is arbitrary, these two criteria will be used to illustrate aspects of dynamic heterogeneity.[66,73–75] The quantity $\tilde{\tau}_{\alpha,p}$ will account for the effect of the long-time part of the relaxation curve in **Figure 3**. As shown in the literature,[66] the long-time decay of the relaxation function is typically associated with the relaxation of the slowest fraction of particles and hence carries information on strong dynamic heterogeneity effects. These long-time decays negligibly contribute to the value $\tilde{\tau}_{\alpha,p'}$ defined at $F_s(\tilde{q}, \tilde{\tau}_{\alpha,p'}) = 1/e$, which instead only weakly accounts for dynamic heterogeneity by considering the most generic particles. We consider $\tilde{\tau}_{\alpha,p'}$ obtained here to be closer to the average hopping time predicted in the accompanying theoretical work,[41] since the theoretical alpha times are computed via a mean-field (most probable time computed) approach and dynamic heterogeneity (in the sense of a distribution of hopping times of dynamic and/or structural origin) is not introduced, although it can be included.[76,77] Differences between how $\tilde{\tau}_{\alpha,p}$ and $\tilde{\tau}_{\alpha,p'}$ depend on $f_{cross}$ do depend on both $T$ and $d/\sigma$, and as we will show can be very small or substantial. We will primarily report $\tilde{\tau}_{\alpha,p}$, considering $\tilde{\tau}_{\alpha,p'}$ only where the comparison is instructive. We can thus extract the key quantities in **Equation 1** directly from molecular simulations, the long-time diffusion $\tilde{D}_p$ and a characteristic hopping or penetrant alpha relaxation time $\tilde{\tau}_{\alpha,p}$, and examine how they are related and affected by the state of the network and penetrant.

We start by considering $\tilde{\tau}_{\alpha,p}$, which we indicated in **Figure 3** and is typically between $\tilde{\tau}_{\alpha,p} = 10^1 - 10^3$ for $d/\sigma = 1.0$ and $\tilde{\tau}_{\alpha,p} = 10^1 - 10^4$ for $d/\sigma = 2.0$. Our goal is not to determine any precise dependence on $d/\sigma$, but rather to explore the differences between relatively 'small' and 'large' molecular penetrants in the context of experimental systems. We plot the inverse penetrant mean alpha times $1/\tilde{\tau}_{\alpha,p}$ or $1/\tilde{\tau}_{\alpha,p'}$ as a function of $T_g(f_{cross})/T$ for several different temperatures $T$ and crosslink fractions $f_{cross}$ in **Figure 4**. We emphasize that the abscissa combines two parameters we consider in our model; the temperature $T$ is plotted in different colors in all panels of **Figure 4a** and **b** for $d/\sigma = 1.0$ and $d/\sigma = 2.0$, respectively, and is directly set in our simulations. However, we note that $T_g/T$ has multiple values for a given $T$ due to the dependence of $T_g$ on $f_{cross}$.[39] For the **Figure 4a** inset, we replot the data to show the corresponding plot of $1/\tilde{\tau}_{\alpha,p}$ versus $T_g/T$, but grouping data at fixed $f_{cross}$ instead of fixed $T$.

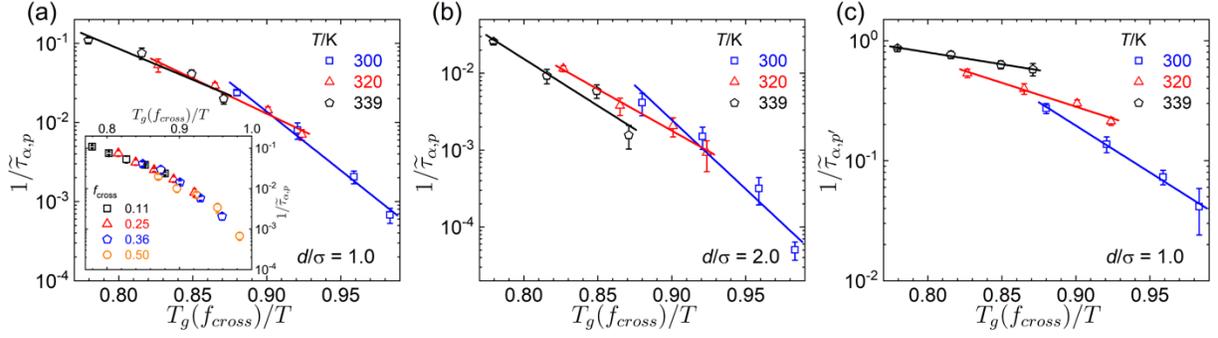

**Figure 4.** Inverse penetrant alpha time $1/\tilde{\tau}_{\alpha,p}$ calculated from simulations as a function of $T_g(f_{cross})/T$ for several temperatures $T$ and penetrant sizes (a) $d/\sigma = 1.0$ and (b) $d/\sigma = 2.0$. In (a) there is a noticeable collapse to a single curve, which we illustrate more clearly in the inset by grouping the data with respect to $f_{cross}$. This collapse is no longer observed in (b). We also plot the shorter-time criteria for the penetrant alpha time $1/\tilde{\tau}_{\alpha,p'}$ versus $T_g(f_{cross})/T$ in (c) for $/\sigma = 1.0$ to demonstrate that the coupling of penetrant hopping to crosslink density is noticeably weaker when the long-time part of the intermediate scattering function is no longer included. The dimensionless slopes of these plots are calculated and included in **Table 1**.

**Figures 4a** and **b** show that at a fixed $T$, the inverse penetrant alpha relaxation time $1/\tilde{\tau}_{\alpha,p}$ behaves exponentially with $T_g(f_{cross})$, exhibiting Arrhenius-like behavior for both penetrant sizes. As the temperature increases, the absolute value of the slope of this curve decreases, demonstrating stronger effects of glass physics at low temperatures in relation to crosslink-induced $T_g$ changes. This underscores the importance of the network segmental dynamics on penetrant diffusion, and can be quantified by a change in the dimensionless slope of $\log \tilde{\tau}_{\alpha,p}$ versus $T_g/T$. To highlight how the choice of criterion $F_s(\tilde{q}, \tilde{\tau}_{\alpha,p}) = 0.1$ affects our results, we plot in **Figure 4c** how **Figure 4a** would be changed upon choosing the alternative criterion, $F_s(\tilde{q}, \tilde{\tau}_{\alpha,p'}) = 1/e$. In this case, similar trends are observed (i.e. an increasing dimensionless slope with decreasing temperature), however the magnitudes of the dimensionless slopes are significantly smaller. We speculate that this is due to the neglect of longer time-scale processes that are apparent in the extended tails of the relaxation functions seen in **Figure 3**, meaning that the 1/e criterion does not fully capture important aspects of dynamic heterogeneity effects that are accounted for with the 0.1 criterion. For both criteria and penetrant sizes $d/\sigma$, the values for the slopes of $\log \tilde{\tau}_{\alpha,p}$ versus $T_g/T$ are calculated and tabulated in **Table 1** for different values of $T$. We note that for both penetrant sizes, these dimensionless slopes exhibit similar trends to the predictions from theory in both our companion theory paper and prior work,[40,41] showing a modest decrease in the magnitude of the slope with increasing temperature for both $\tilde{\tau}_{\alpha,p}$ and $\tilde{\tau}_{\alpha,p'}$. However, the magnitudes of the slopes are significantly larger for $\tilde{\tau}_{\alpha,p}$ than the slopes for $\tilde{\tau}_{\alpha,p'}$, with the latter values in parenthesis in **Table 1**. We attribute this to the inclusion of the longer-time component of the ISF in $\tilde{\tau}_{\alpha,p}$, which indicate that penetrant hopping is influenced by dynamic heterogeneity near $T_g$ due to the slower relaxation of a significant fraction of particles.[66] To contrast, the longer-time decays of the ISF negligibly affect $\tilde{\tau}_{\alpha,p'}$, leading to a weaker apparent coupling between the near-$T_g$ dynamics of the network and the penetrant hopping.

**Table 1.** Dimensionless slope for $\log \tilde{\tau}_{\alpha,p}$ or $\widetilde{D}_p^{-1}$ versus $T_g/T$ from simulations at several temperatures ($T = 300$, 320, and 339K) and penetrant sizes ($d/\sigma = 1$ and 2). Both criteria $\tilde{\tau}_{\alpha,p}$ and $\tilde{\tau}_{\alpha,p'}$ (in parentheses) are considered.

| T/K | $d/\sigma$ | 300 | 320 | 339 |
|---|---|---|---|---|
| $\log(\tilde{\tau}_{\alpha,p})$ vs $T_g/T$ | 1.0 | 15.57 (7.79) | 8.59 (3.98) | 8.01 (1.97) |
| $\log(1/D_p)$ vs $T_g/T$ | 1.0 | 9.34 | 7.48 | 5.79 |
| $\log(\tilde{\tau}_{\alpha,p})$ vs $T_g/T$ | 2.0 | 18.13 (16.10) | 12.86 (8.78) | 14.80 (9.60) |
| $\log(1/D_p)$ vs $T_g/T$ | 2.0 | 17.47 | 19.27 | 18.12 |

The two different penetrant sizes in **Figure 4a** and **4b** exhibit similar trends, including nearly the same overall decrease of relaxation rate by a little over two decades (over all values of $T$ and $f_{cross}$). This supports the robustness of our analysis and conclusions drawn. However, there are subtle but important differences. At a fixed value of $T_g(f_{cross})/T$ (i.e. at a fixed supercooling degree), we observe that the penetrant alpha time is roughly constant with decreasing temperature for $d/\sigma = 1.0$. This lower temperature also corresponds to a lower value of $T_g(f_{cross})$ and hence a lower degree of crosslinking $f_{cross}$. This means that the degree of effective supercooling as encoded in the ratio $T_g(f_{cross})/T$ largely dictates penetrant hopping, with the absolute temperature $T$ playing a similar role in affecting transport as the effective $T_g$. In this case, several effects – changing $T_g$, absolute $T$, and crosslinking ($f_{cross}$) – tend to compensate resulting in a near collapse of the $1/\tilde{\tau}_{\alpha,p}$ versus $T_g/T$ data. This is highlighted in the inset of **Figure 4a**, which is the same data as in the main frame but grouped by $f_{cross}$ instead of $T$, and more clearly shows how well this data collapses to a single curve. A similar collapse of $1/\tilde{\tau}_{\alpha,p}$ versus $T_g/T$ is also observed in the companion theory paper.[41] **Figure 4b** shows that this balance no longer holds for larger penetrants $d/\sigma = 2.0$, with lower $T$ values exhibiting faster dynamics (i.e. higher $1/\tilde{\tau}_{\alpha,p}$) at a given degree of supercooling $T_g(f_{cross})/T$ due to the corresponding decrease in the $f_{cross}$. We suggest that, for $d/\sigma = 2.0$, the importance of dynamic heterogeneity is less pronounced relative to that of $d/\sigma = 1.0$ penetrants due to a greater degree of self-averaging (in time and space) of the distribution of matrix relaxation times by a larger penetrant. Additionally, the effect of decreasing crosslinking dominates more than the effect of decreasing temperature, resulting in a higher value of $1/\tilde{\tau}_{\alpha,p}$. This contrasts with **Figure 4c**, where $1/\tilde{\tau}_{\alpha,p'}$ increases with temperature at a fixed $T_g(f_{cross})/T$; here, dynamic heterogeneity becomes negligible, and the effects of crosslinking become even less pronounced than the absolute temperature.

    This same competition is apparent in the diffusion constant $\tilde{D}_p$, which we plot in **Figure 5a** and **b** for $d/\sigma = 1.0$ and $d/\sigma = 2.0$, respectively, and at the same conditions plotted in **Figure 4a** and **b**. As discussed above, the diffusion constant is defined at longer time and length scales than the length scale of the crosslinking mesh size (see **Figures 2** and **S1**). This quantity is

influenced by dynamic heterogeneity effects at lower temperatures, where glassy dynamics becomes important, but is also affected by the presence of crosslinkers. We thus expect its behavior to be similar to $1/\tilde{\tau}_{\alpha,p}$ instead of $1/\tilde{\tau}_{\alpha,p'}$, as confirmed below. Like $1/\tilde{\tau}_{\alpha,p}$, $\widetilde{D}_p$ exhibits Arrhenius-like behavior as defined by a straight line on a $\log \widetilde{D}_p$ versus $T_g/T$ plot. The primary difference is in the relative magnitude change in $D_p$ at different degrees of supercooling $T_g/T$. $\widetilde{D}_p$ still exhibits a near-collapse for $d/\sigma = 1.0$, however generally decreases more slowly than $1/\tilde{\tau}_{\alpha,p}$ in **Figure 4a**.

For the larger penetrant size $d/\sigma = 2.0$, the situation is more complicated. In this case, high temperatures $T$ in **Figure 5b** show a more dramatic decrease in $\widetilde{D}_p$ with $T_g/T$ than that of $1/\tilde{\tau}_{\alpha,p}$ in **Figure 4b**, but a similar decrease at lower temperatures. These effects are quantified by calculating the dimensionless slope of the $-\log \widetilde{D}_p$ versus $T_g/T$ relationship, which we include in **Table 1** for both penetrant sizes. Here, the $\widetilde{D}_p$ slopes are significantly smaller than the $1/\tilde{\tau}_{\alpha,p}$ slopes for $d/\sigma = 1.0$, but are similar or larger than the $1/\tilde{\tau}_{\alpha,p}$ for $d/\sigma = 2.0$. The latter behavior for $d/\sigma = 2.0$ is qualitatively consistent with the results in our companion theory paper,[41] which finds that the diffusion constant $\widetilde{D}_p$ changes with supercooling $T_g/T$ are similar to, or more pronounced than, the corresponding inverse $\tilde{\tau}_{\alpha,p}$. For $d/\sigma = 1.0$, however, the theory predicts that the $\widetilde{D}_p$ slopes are similar to the inverse $\tilde{\tau}_{\alpha,p}$ slopes, in contrast to the simulation observation that the $\widetilde{D}_p$ slopes are much smaller. We attribute this disparity between simulation and theory to the larger importance of dynamic heterogeneity for smaller penetrants in our simulations,[66] because the theory only makes predictions for a mean (in a Kramers first passage time sense) penetrant alpha time that is closer to the value of $\tilde{\tau}_{\alpha,p'}$. Indeed, for $d/\sigma = 1.0$ the relative slopes for $\widetilde{D}_p$ and $1/\tilde{\tau}_{\alpha,p'}$ calculated using this criterion (in parentheses in **Table 1**) are much more consistent with the theoretical predictions.[41]

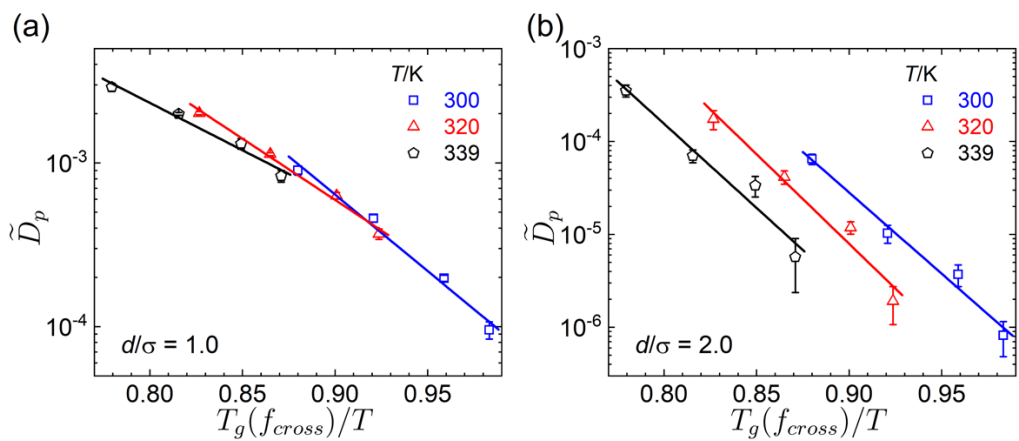

**Figure 5.** (a) Penetrant diffusion constant $\tilde{D}_p$ as a function of $T_g(f_{cross})/T$ for several temperatures $T$ and penetrant sizes (a) $d/\sigma = 1.0$ and (b) $d/\sigma = 2.0$. These are the same conditions as in **Figure 4a** and **b**, and similarly there is a collapse in (a) that is no longer present in (b). The dimensionless slopes of these plots are calculated and included in **Table 1**.

To contextualize the relationship between the diffusion constant $\tilde{D}_p$, inverse penetrant alpha time $1/\tilde{\tau}_{\alpha,p}$, temperature $T$, and particle size ratio $d/\sigma$, we plot both $\tilde{D}_p$ and $1/\tilde{\tau}_{\alpha,p}$ as a function of $T_g/T$ in **Figure 6** for two temperatures each. We have used a multiplier $A$ to align these data sets at the leftmost points, to compare how the effect of changing the crosslinking fraction $f_{cross}$ (and concomitantly the $T_g$) on $\tilde{D}_p$ versus $1/\tilde{\tau}_{\alpha,p}$. For small particles ($d/\sigma = 1.0$), the value of $\tilde{D}_p$ (solid symbols) exhibits a weaker decrease with the degree of supercooling $T_g/T$ than $1/\tilde{\tau}_{\alpha,p}$ (open symbols) for both temperatures. We believe that this is again a dynamic heterogeneity effect, which may be related to the well-known breakdown of the Stokes-Einstein relation in one-component glass forming liquids,[66–73] where the diffusion constant is understood to be dominated by the subpopulation of particles undergoing rapid transport and is thus less dependent on $T_g/T$ than the particle hopping time that characterizes a relaxation process.[66] To contrast, for larger particles ($d/\sigma = 2.0$) the change in the diffusion constant $\widetilde{D}_p$ tracks the change in $1/\tilde{\tau}_{\alpha,p}$ with increasing $f_{cross}$ at low temperatures ($T = 300$K) and even decreases more than $1/\tilde{\tau}_{\alpha,p}$ with increasing $f_{cross}$ at higher temperatures ($T = 339$K). This indicates that there is an additional mechanism slowing down diffusive penetrant motion at length scales longer than what is probed in the intermediate scattering function (the local penetrant cage corresponding to $1/\tilde{\tau}_{\alpha,p}$), which we attribute to the effect of mesh confinement. A similar comparison between $\widetilde{D}_p$ and $1/\tilde{\tau}_{\alpha,p'}$ (not shown) for both sizes demonstrates that $\widetilde{D}_p$ exhibits a stronger decrease with $T_g/T$ than $1/\tilde{\tau}_{\alpha,p'}$ for all temperatures studied. This does not relate to the normal Stokes-Einstein breakdown, because it only weakly accounts for dynamic heterogeneity, but does further suggest the importance of the mesh confinement effect on the diffusion constant.

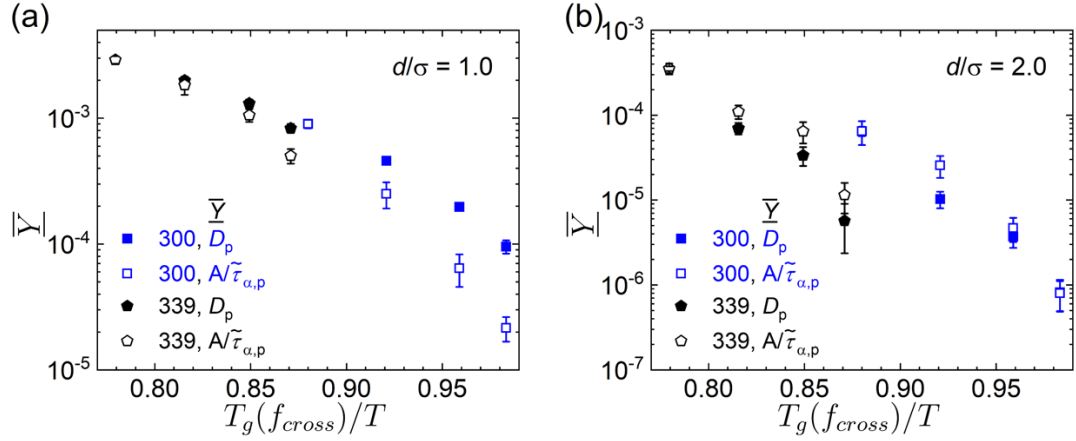

**Figure 6.** Comparison of the penetrant diffusion constant $\widetilde{D}_p$ and inverse penetrant alpha time $1/\tilde{\tau}_{\alpha,\text{p}}$ as a function of $T_g(f_{cross})/T$ at both high (339K) and a low (300K) temperatures for (a) $d/\sigma = 1.0$ and (b) $d/\sigma = 2.0$. Note that $1/\tilde{\tau}_{\alpha,\text{p}}$ is vertically shifted down by a factor (a) $A = 3.89 \times 10^{-2}$ and (b) $A = 2.53 \times 10^{-2}$ such that the leftmost point of alpha time $1/\tilde{\tau}_{\alpha,\text{p}}$ at a given temperature coincides with the corresponding $\widetilde{D}_p$ data point for better comparison. In (a), this comparison shows that $\widetilde{D}_p$ exhibits a weaker dependence than $1/\tilde{\tau}_{\alpha,\text{p}}$ with $T_g(f_{cross})/T$ for all temperatures, while in (b) $\widetilde{D}_p$ exhibits a stronger dependence than $1/\tilde{\tau}_{\alpha,\text{p}}$ at high temperatures.

To isolate the role of mesh confinement, we refer to the form of **Equation 1** that writes the diffusion constant as the product of $1/\tilde{\tau}_{\alpha,p}$ and a mesh confinement factor $X$. The form predicted for $X$ is model-dependent,[27,28,30] but we can directly evaluate this quantity from simulation by taking the product $X = \widetilde{D}_p \tilde{\tau}_{\alpha,p}$. We plot $\widetilde{D}_p \tilde{\tau}_{\alpha,p}$ in **Figure 7** for both penetrant sizes as a function of $T_g/T$ and $T$, in analogy with **Figures 4** and **5**. **Figure 7a** plots this product for small penetrants ($d/\sigma = 1.0$), and as expected from **Figures 4a** and **5a** the data roughly collapses to a single curve that increases monotonically with the degree of supercooling $T_g/T$. This trend is expected for one-component glass-forming liquids due to alpha process dynamic heterogeneity,[66] so in this case the network primarily affects penetrant transport through the dependence of $T_g$ on $f_{cross}$. Notably, this monotonic increase in $\widetilde{D}_p \tilde{\tau}_{\alpha,p'}$ is no longer seen when the 1/e criterion is considered (see inset to **Figure 7a**), demonstrating that the longer time decay in the ISF should be responsible for the behavior in the main panel of **Figure 7a**. To contrast, **Figure 7b** plots the product $\widetilde{D}_p \tilde{\tau}_{\alpha,p}$ for larger penetrants ($d/\sigma = 2.0$), and they exhibit a significant *decrease* with $T_g/T$ for each of the higher temperatures $T = 339$ and 320K. For a constant temperature, the increase with $T_g/T$ corresponds to an increase in both $f_{cross}$ and $C$. We can thus attribute this dependence to the effect of mesh confinement (and also the reduced importance of dynamic heterogeneity since larger particles tend to average over it per discussions above), as this decrease of $\widetilde{D}_p \tilde{\tau}_{\alpha,p} = X$ with $C$ is predicted by both Cai, *et al.*[27] and Dell and Schweizer.[30] However, this trend is less noticeable for the lowest temperature $T = 300$ K, where within error $\widetilde{D}_p \tilde{\tau}_{\alpha,p}$ remains roughly constant (slightly nonmonotonic) because at lower temperatures the dynamic heterogeneity effect becomes more important. The expected increase in $\widetilde{D}_p \tilde{\tau}_{\alpha,p}$ for glassy melts due to dynamic heterogeneity appears to cancel out any mesh confinement effects, indicating that the local caging and mesh confinement both contribute to penetrant dynamics in this limit.

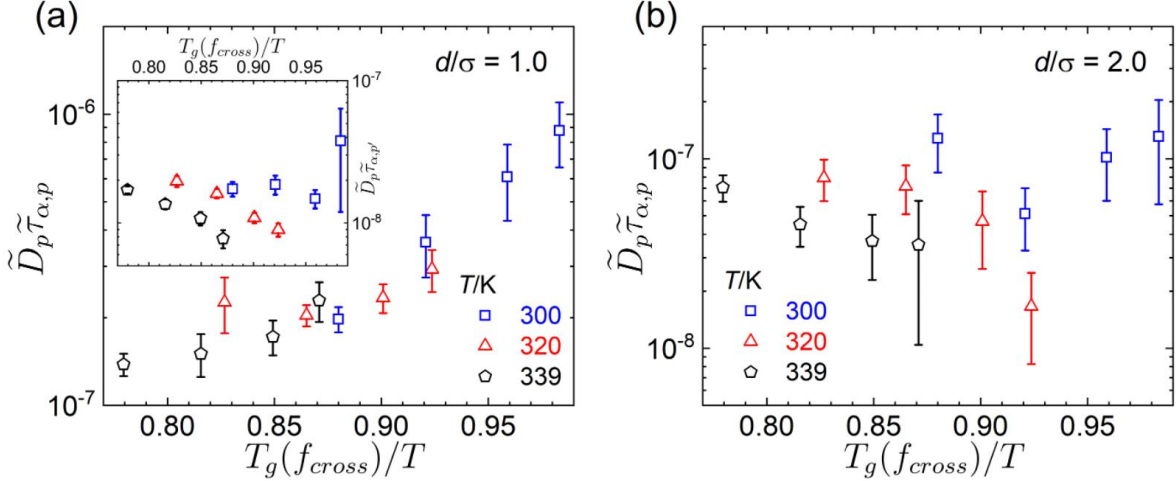

**Figure 7.** The product $\widetilde{D}_p \widetilde{\tau}_{\alpha,p}$ as a function of $T_g(f_{cross})/T$ for several temperatures $T$ for (a) $d/\sigma = 1.0$ and (b) $d/\sigma = 2.0$ with the 0.1 ISF decay criterion. For (a) there is a collapse to an increasing curve that reflects the breakdown of Stokes-Einstein behavior typically observed in fragile glass forming liquids. The inset shows the same plot, but using the 1/e criterion. For (b) at high temperatures, each temperature exhibits a monotonic decrease in the product $\widetilde{D}_p \widetilde{\tau}_{\alpha,p}$ that we attribute to the effect of entropic mesh confinement. At low temperatures ($T = 300K$) and $d/\sigma = 2.0$, the increase due to dynamic heterogeneity appears to counteract the effect of mesh confinement, leading to a roughly constant value of $\widetilde{D}_p \widetilde{\tau}_{\alpha,p}$.

*Network and Penetrant Dynamic Decoupling*

To quantify penetrant dynamics we determined both the rate of local hopping via $1/\widetilde{\tau}_{\alpha,p}$ and the long-time diffusion via $\widetilde{D}_p$. Our results demonstrated that (1) both processes were primarily governed by local network segmental motion (i.e. glass physics) at low temperatures, yet (2) crosslinks nevertheless had a more pronounced effect on $\widetilde{D}_p$ for larger penetrants ($d/\sigma = 2.0$ versus $d/\sigma = 1.0$ in **Table 1**) or for situations that dynamic heterogeneity is not important. To understand these two observations, we quantify the coupling between the penetrant alpha time $\widetilde{\tau}_{\alpha,p}$, and the network alpha relaxation time $\widetilde{\tau}_{\alpha,K}$, where the latter is obtained from the autocorrelation function of the Kuhn monomer vector per our prior work as briefly described in the methods section.[39] In this case, we use the short-time criterion $\tilde{\tau}_{\alpha,p'}$ due to its better consistency with both the theoretical predictions,[41] the weaker effect of dynamic heterogeneity, and to match the $1/e$ criterion used in calculating $\tilde{\tau}_{\alpha,K}$ previously.[39] The ratio between these two properties $\tilde{\tau}_{\alpha,p'}/\tilde{\tau}_{\alpha,K}$ is plotted as a function of $T_g/T$ in **Figure 8** for both penetrant sizes and all temperatures $T$ studied; this comparison has previously demonstrated to be useful in the self-consistent cooperative hopping (SCCH) theory of hard sphere mixtures and dilute penetrants in polymer melts,[24,65,78] and quantifies the degree of 'dynamic coupling/decoupling'. In these previous theoretical papers, the ratio $\tilde{\tau}_{\alpha,p'}/\tilde{\tau}_{\alpha,K}$ was studied as a function of 'packing fraction', which is qualitatively analogous to inverse temperature.[65] The theory predicted that the ratio $\tilde{\tau}_{\alpha,p'}/\tilde{\tau}_{\alpha,K}$ (1) initially increases slightly as the temperature is lowered from high temperature, and then (2) sharply decreases in a penetrant size-dependent manner (stronger decrease for smaller penetrants) as temperature is further lowered towards the glassy state. This indicates a strong decoupling between penetrant and matrix dynamics in the supercooled regime.

In **Figure 8**, both penetrants show only a modest dependence of $\tilde{\tau}_{\alpha,p'}/\tilde{\tau}_{\alpha,K}$ on $T_g/T$, spanning only roughly a decade in the simulation-accessible weakly supercooled regime. We do not observe the characteristic decrease in $\tilde{\tau}_{\alpha,p'}/\tilde{\tau}_{\alpha,K}$ theoretically predicted for the deeply supercooled regime, which is expected to be computationally inaccessible, though our trends are consistent with the weak changes in $\tilde{\tau}_{\alpha,p'}/\tilde{\tau}_{\alpha,K}$ predicted by the theory in the polymer alpha relaxation time range probed in our simulation.[41] Importantly, the relative magnitude of these ratios is instructive. We find $\tilde{\tau}_{\alpha,p'}/\tilde{\tau}_{\alpha,K}$ for the $d/\sigma = 2.0$ penetrant is at least an order of magnitude larger than that for the $d/\sigma = 1.0$ penetrant, indicating that it is more strongly coupled to the surrounding matrix as generically expected and as predicted by SCCH theory.[24,41,65,78] This helps explain our results for $1/\tilde{\tau}_{\alpha,p}$ and $\widetilde{D}_p$ in the previous section; in those results, there was a stronger dependence of $\widetilde{D}_p$ for the larger penetrants on $f_{cross}$ even at lower temperatures where mesh confinement is not dominant (see **Table 1**). We can now attribute this to the stronger matrix-penetrant coupling for larger penetrants, rather than enhanced mesh confinement contribution.

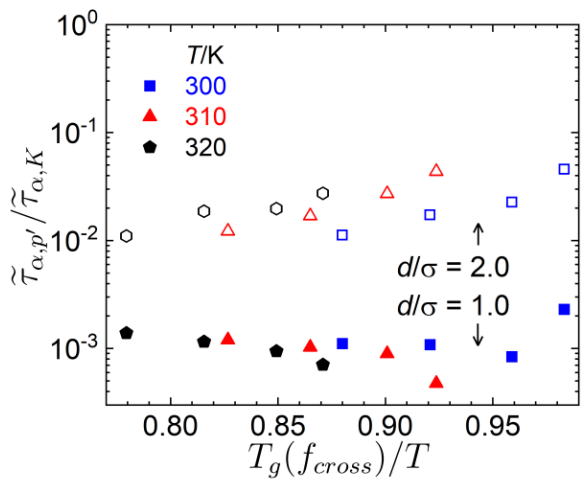

**Figure 8.** Degree of decoupling between penetrant and Kuhn segment alpha relaxation times quantified by the ratio $\tilde{\tau}_{\alpha,p'}/\tilde{\tau}_{\alpha,K}$, as a function of $T_g(f_{cross})/T$ for several temperatures and both $d/\sigma = 1.0$ and $d/\sigma = 2.0$. While the trends with $T_g(f_{cross})/T$ are quite weak, the larger penetrant $d/\sigma = 2.0$ couples significantly more strongly to the segmental dynamics of the surrounding network, as evident in the order-of-magnitude larger value of $\tilde{\tau}_{\alpha,p'}/\tilde{\tau}_{\alpha,K}$ when compared to the smaller penetrant $d/\sigma = 1.0$.

*Degeneracy of Mesh Size Effects Due to Confinement and Glassy Dynamics*

From a molecular viewpoint, we only see significant mesh confinement effects on penetrant diffusion in the limit of large particles and high temperatures, while for small molecules or low temperatures we find that crosslinking dominantly impacts penetrant diffusion primarily through its effect on the local segmental relaxation of the polymer matrix. However, even in this latter limit we can demonstrate results *empirically* consistent with aspects of **Equation 1** if we plot our data versus the confinement ratio $C$ rather than the extent of supercooling $T_g/T$. We consider two candidate models for $X$, by plotting $C\widetilde{D}_p$ versus $C^2$ in **Figure 9a** based on the predictions of Cai, *et al.*[27] and $\widetilde{D}_p$ versus $C$ in **Figure 9b** based on the predictions of Dell and Schweizer.[30] Despite already demonstrating that mesh confinement does *not* play a significant role for small penetrants, these plots exhibit remarkably good agreement with the functional forms predicted by both models as indicated by the nearly linear trends in these semi-log plots. However, there is a non-negligible change in slope in both plots of **Figure 9** as the temperature $T$ is changed, which is not consistent with either entropic mesh confinement model.[27,29,30] As per **Equation 1**, these models that only have a temperature dependence in the prefactor $1/\tau_{\alpha,p}$ (plotted also versus $C$ from simulation in **Figure S2)** and not in the exponent of the entropic mesh confinement factor $X$. This key observation, which is also shown theoretically in the companion paper,[41] allows the effects of crosslinking on the penetrant relaxation time to be distinguished from the effects of crosslinking due to mesh confinement, in the absence of simulation determination of $1/\tau_{\alpha,p}$, though this distinguishing feature may become weak as in the case of larger penetrants as is shown in the supporting information ($d/\sigma = 2.0$, **Figure S3**).

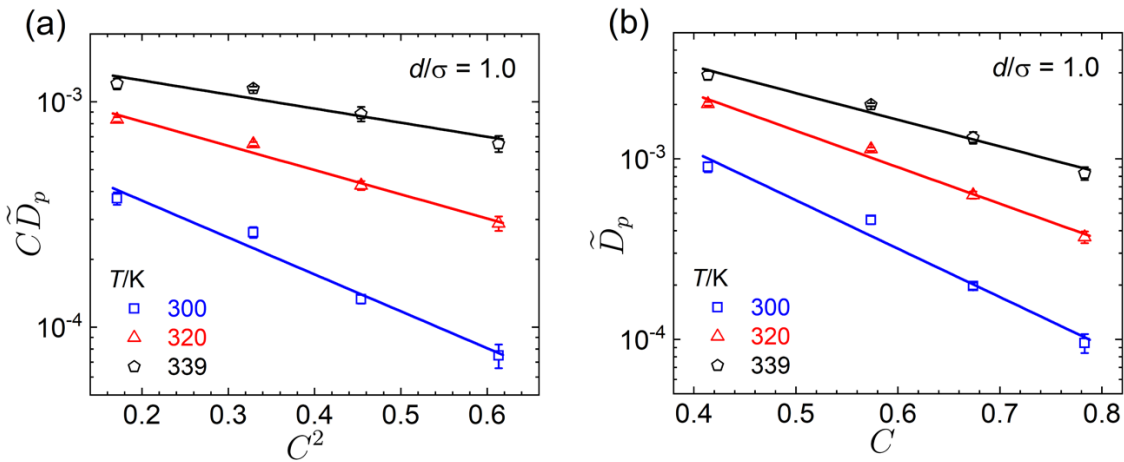

**Figure 9.** The penetrant diffusion constant $\widetilde{D}_p$ plotted with respect to the confinement ratio $C$ as suggested by the forms of various literature predictions for mesh confinement effects. (a) Cai, *et al.*[23] predicts that $\widetilde{D}_p \propto C^{-1} \exp(-C^2)$ and (b) Dell and Schweizer[26] predict that $\widetilde{D}_p \propto \exp(-bC)$. Both ways of plotting this data appear linear, even though mesh confinement does not play a major role in penetrant transport for the $d/\sigma = 1.0$ case plotted here. The main indication that the simulations do *not* agree with the idea that entropic mesh confinement is dominant lie in the temperature-dependent slopes, which for mesh confinement should not change with temperature.

**Conclusion**

In this paper, we used simulation to understand the role of glassy segmental scale dynamics versus entropic mesh confinement on the diffusion and alpha relaxation of molecular penetrants in permanently crosslinked polymer networks. We consider a specific form for the diffusion constant, $D_p \sim (d^2/\tau_{\alpha,p})X$, that allows us to isolate the effect of network mesh confinement $X$ through calculating and studying both the diffusion constant $\widetilde{D}_p$ and the penetrant hopping or alpha relaxation time $\widetilde{\tau}_{\alpha,p}$. Both quantities exhibit Arrhenius-like trends with the degree of effective supercooling $T_g/T$ at a fixed temperature, due to the dependence of $T_g$ on the extent of crosslinking. This strong coupling between penetrant hopping and crosslinking-dependent glassy dynamics is consistent with theoretical predictions in the companion paper,[41] but also exhibits significant breakdown in the Stokes-Einstein relationship if strong dynamic heterogeneity effects[66,73–75] are apparent in the hopping dynamics of the smaller molecular penetrants as the system is supercooled to larger values of $T_g/T$.[66] For small penetrants, the effect of crosslinking on $T_g$ is demonstrated to dominate diffusive motion, with a collapse to a single dependence of both $\widetilde{\tau}_{\alpha,p}$ and $\widetilde{D}_p$ on $T_g/T$. This collapse is no longer apparent for larger penetrants, where we show by plotting the product $\widetilde{D}_p \widetilde{\tau}_{\alpha,p}$ as a function of $T_g/T$ that we see the signatures of mesh confinement. While we find that crosslinking affects molecular penetrant transport through *both* its effect on $T_g$ and through mesh confinement, the former appears to be the dominant mechanism for our simulations and we expect it will also be the dominant for deeply supercooled experiments. We can also show, however, that the diffusion constant $\widetilde{D}_p$ still follows trends predicted in the literature for mesh confinement,[27,28,30] due to a degeneracy of how crosslink fraction changes with $T_g$ and mesh size as predicted by SCCH theory.[41] This underscores the difficulty of drawing conclusions on transport mechanisms from diffusion data alone.

The importance of activated glassy dynamics on the diffusive motion of small molecular penetrants in polymer networks suggests directions for future study, and ways in which the selective transport of these species could be engineered into polymer materials. In particular, we anticipate that the role of molecular shape and penetrant-matrix interactions may complicate the relations we show between entropic mesh confinement and segmental relaxation. For penetrants that are highly non-spherical, their size will now be related to two length-scales that may be more or less affected by entropic mesh confinement, and will possess anisotropic interactions with the surrounding matrix that may affect the penetrant alpha time. While related questions have been theoretical studied by some of the authors,[24,26] simulation will provide further molecular insight into the role of these various transport effects. Finally, the important role of dynamic heterogeneity on penetrant hopping suggests that there are possibilities for tuning the selectivity of molecular transport in networks with dynamic or exchangeable bonds, which would introduce another competing timescale that has recently been shown experimentally to affect penetrant diffusion.[79]

**Acknowledgement**

This research was supported by the U.S. Department of Energy, Office of Basic Energy Sciences, Division of Materials Sciences and Engineering (Award No. DE-SC0020858), through the Materials Research Laboratory at the University of Illinois at Urbana-Champaign. Helpful discussions with Christopher Evans are gratefully acknowledged.

**Supporting Information Figures**

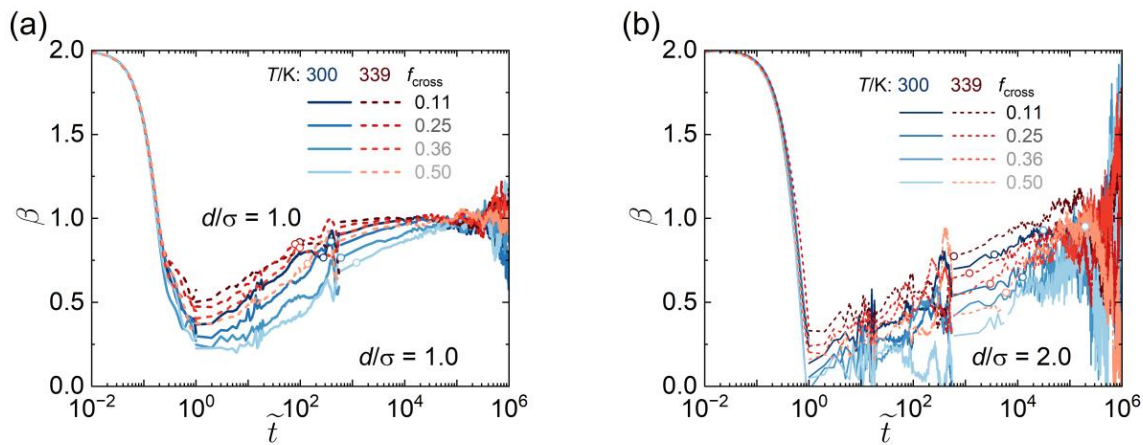

**Figure S1**. Representative plots of the effective exponent $\beta$ for the mean square displacement versus time $\langle \tilde{r}_p^2(\tilde{t}) \rangle \sim \tilde{t}^\beta$ of a penetrant particle with sizes (a) $d/\sigma = 1.0$ and (b) $d/\sigma = 2.0$, for the same values of $T$ and $f_{cross}$ as shown in **Figure 2**. $\beta$ is the slope of **Figure 2**, and both plots show that penetrant diffusion reaches the Fickian diffusion limit of $\beta = 1$ in the long-time region. The open circles denote the length scale of the polymer network mesh size, as described in **Table S1**. Intermediate regions with more noise reflect the need to perform several distinct computational runs with different data output frequency.

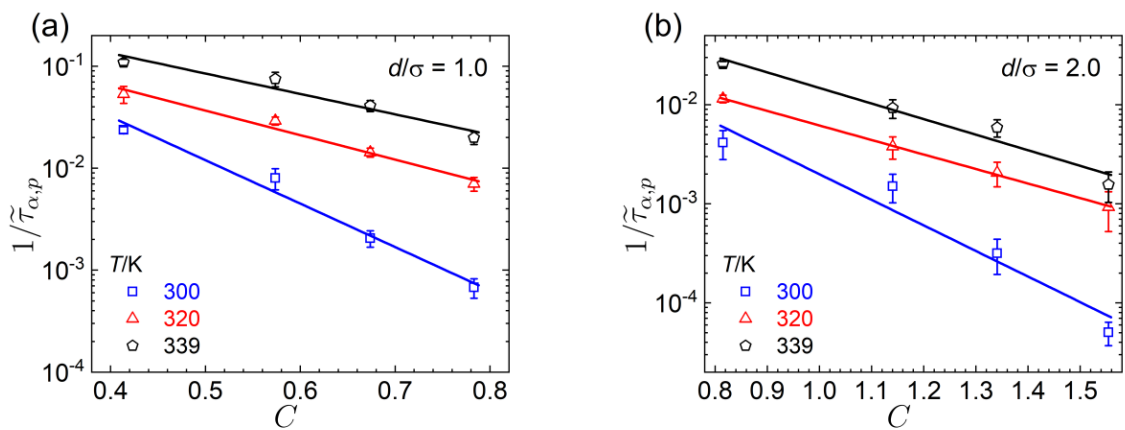

**Figure S2.** The inverse penetrant alpha time $1/\tilde{\tau}_{\alpha,p}$ plotted with respect to the confinement ratio $C$ for penetrant particles with sizes (a) $d/\sigma = 1.0$ and (b) $d/\sigma = 2.0$. For both particle sizes, we find that they are consistent with the trend $\tilde{\tau}_{\alpha,p} \propto \exp(-bC)$. This primarily reflects the local glassy dynamics and *not* mesh confinement effects, and so this trend contributes to the apparent degeneracy between the two mechanisms for crosslink effects on penetrant transport in networks.

**Figure S3:**

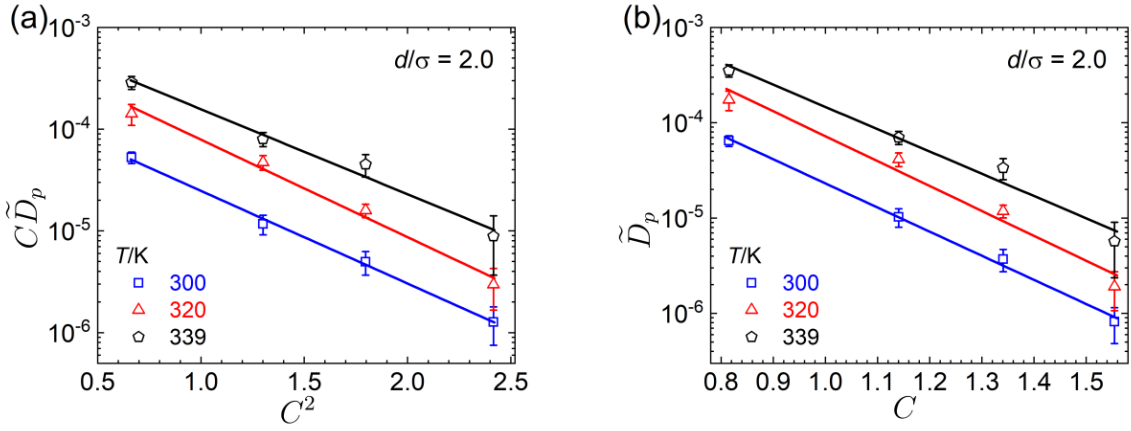

**Figure S3.** The penetrant diffusion constant $\widetilde{D}_p$ plotted with respect to the confinement ratio $C$ as suggested by the forms of various literature predictions for mesh confinement effects. (a) Cai, *et al.*[23] predicts that $\widetilde{D}_p \propto C^{-1} \exp(-C^2)$ and (b) Dell and Schweizer[26] predict that $\widetilde{D}_p \propto \exp(-bC)$. Both ways of plotting this data appear linear, even though mesh confinement does not play the sole role in penetrant transport for the $d/\sigma = 2.0$ case shown here.

**Table S1.** Mesh size (in unit of $\sigma$) of networks at various $f_{\text{cross}}$.

| $f_{\text{cross}}$ | 0.11 | 0.25 | 0.36 | 0.50 |
|---|---|---|---|---|
| $d/\sigma = 1.0$ | 2.43 | 1.75 | 1.48 | 1.28 |
| $d/\sigma = 2.0$ | 2.45 | 1.76 | 1.49 | 1.29 |